\documentclass[%
reprint, 
 amsmath,amssymb,
 aps,
]{revtex4-2}

\usepackage{graphicx}
\usepackage{dcolumn}
\usepackage{bm}

\usepackage[utf8]{inputenc}
\usepackage{float} 
\usepackage{makecell}
\usepackage{hyperref}
\usepackage{mathrsfs}
\usepackage{graphicx} 
\usepackage[markupon]{qnnmarkup}
\usepackage[markupon]{qnnmarkup}
\usepackage{qnnmarkup}
\usepackage{siunitx}
\usepackage{amsmath,amssymb}
\usepackage{setspace}
\newcommand{\Var}[1]{\text{Var}\left(#1\right)}

\newcommand{\mean}[1]{\left\langle#1\right\rangle}
\newcommand{\Neg}{{N_e^{(g)}}}
\newcommand{\Np}{{N_e^{(p)}}}
\newcommand{\Nd}{{N_e^{(d)}}}
\newcommand{\nd}{{n_e^{(d)}}}
\newcommand{\Ng}{{N_\gamma^{(d)}}}
\renewcommand{\ng}{{n_\gamma^{(d)}}}
\newcommand{\Ge}{{\Gamma_e^{(g)}}}
\newcommand{\Gedc}{{\Gamma_e^{(dc)}}}
\newcommand{\Gebg}{{\Gamma_e^{(bg)}}}
\newcommand{\Ggdc}{{\Gamma_\gamma^{(bg)}}}
\newcommand{\dose}{d}
\newcommand{\FRV}[2]{\text{FRV}\left(#1\middle|#2\right)}

\begin{document}

\preprint{Koppell/Electron Heralding}

\title{Analysis and Applications of a Heralded Electron Source}

\author{Stewart A. Koppell}%
 \email{skoppel2@jh.edu}
 \affiliation{Research Laboratory of Electronics, Massachusetts Institute of Technology, Cambridge, Massachusetts 02139, USA}
 
\author{John W. Simonaitis}
 \affiliation{Research Laboratory of Electronics, Massachusetts Institute of Technology, Cambridge, Massachusetts 02139, USA}

\author{Maurice A. R. Krielaart}
 \affiliation{Research Laboratory of Electronics, Massachusetts Institute of Technology, Cambridge, Massachusetts 02139, USA}
 
 \author{William P. Putnam}
 \affiliation{%
 Department of Electrical and Computer Engineering. University of California, Davis, Davis, California 95616, USA
}%

\author{Karl K. Berggren}
 \affiliation{Research Laboratory of Electronics, Massachusetts Institute of Technology, Cambridge, Massachusetts 02139, USA}

\author{Phillip D. Keathley}
 \affiliation{Research Laboratory of Electronics, Massachusetts Institute of Technology, Cambridge, Massachusetts 02139, USA}

\date{\today}

\begin{abstract}
    We analytically describe the noise properties of a heralded electron source made from a standard electron gun, a weak photonic coupler, a single photon counter, and an electron energy filter. We argue the traditional heralding figure of merit, the Klyshko efficiency, is an insufficient statistic for characterizing performance in dose-control and dose-limited applications. Instead, we describe the sub-Poissonian statistics of the source using the fractional reduction in variance and the fractional increase in Fisher Information. Using these figures of merit, we discuss the engineering requirements for efficient heralding and evaluate potential applications using simple models of electron lithography, bright-field scanning transmission electron microscopy (BFSTEM), and scanning electron microscopy (SEM). We find that the advantage in each of these applications is situational, but potentially significant: dynamic control of the trade-off between write speed and shot noise in electron lithography; an order of magnitude dose reduction in BFSTEM for thin samples (e.g. 2D materials); and a doubling of dose efficiency for wall-steepness estimation in SEM. 
\end{abstract}

\maketitle

\textit{Introduction} - Heralded photon sources are an elementary resource in quantum optics. Since their first demonstration in 1977 \cite{kimble1977photon}, they have found wide-ranging application: from their use in quantum key distribution \cite{bennett2014quantum}, simulation \cite{o2007optical}, and sensing~\cite{matthews2016towards}, to defining the candela \cite{cheung2007quantum}.  Recently there has been a surge of interest in applying principles of quantum optics to free electrons. This was sparked in part due to various experimental demonstrations of strong coherent coupling between photons and free electrons in electron microscopes \cite{barwick2009photon, Dahan2020,Henke2021,Shiloh2022,Yang2023}, as well as new theoretical predictions which propose how such interactions could be useful for producing unique optical and free-electron quantum states \cite{Dahan2023,DiGiulio2020,Feist2020}. Most recently, photon-mediated electron heralding in a transmission electron microscope (TEM) was demonstrated for the first time~\cite{Feist2022}. It has also been suggested that heralding could be done using coulomb-correlated electron pairs, which have recently been produced for the first time in a TEM\cite{Haindl2023} and a scanning electron microscope (SEM)\cite{Meier2023}.

In spite of this progress toward heralded electron sources, there has been little discussion of their potential impact. In optical microscopy, shot noise is often an important limiting effect, giving heralded photon sources clear utility \cite{MaganaLoaiza2019}. 
However in high-energy electron microscopy, where detectors can be fast and efficient enough to register nearly every electron, fluctuations in beam current may not contribute significantly to the measurement noise. In addition, the proposed heralding schemes involve filters which would substantially reduce beam current. This raises the question of when exactly electron heralding would improve performance.

Here, we analyze an electron heralding system modeled after the one demonstrated by Feist et al. \cite{Feist2022}. First, we parameterize this system and describe its sub-Poissonian beam statistics in terms of the Klyshko efficiencies. We find that the extra information provided by a heralded source makes it possible to improve the speed, yield, and minimum feature size in electron lithography. With sufficient heralding efficiency, we find an order-of-magnitude increase in write speeds is possible.

We then use the Fisher information formalism to quantify the increase in information available for various modalities of quantitative electron microscopy, and we identify two limits (high dark counts and low contrast) where signal-to-noise ratio (SNR) enhancement is possible for dose-limited imaging. We argue there is situational improvement to SNR for high energy electron microscope applications like scanning transmission electron microscopy (STEM), but not for electron-energy-loss spectroscopy (EELS), or energy-dispersive X-ray spectroscopy (EDS). A heralded source could also be a solution for dose-calibrated in-situ electron microscopy and low background cathodoluminescence (CL) imaging. In addition, we find heralding can improve quantitative SEM when the dose is limited by damage or charging. We calculate the error in surface tilt estimation of a steep-walled feature, finding that heralding can reduce the dose required to reach a prescribed level of measurement error by a factor of more than 2.

\begin{figure}
    \centering
    \includegraphics[width=0.48 \textwidth]{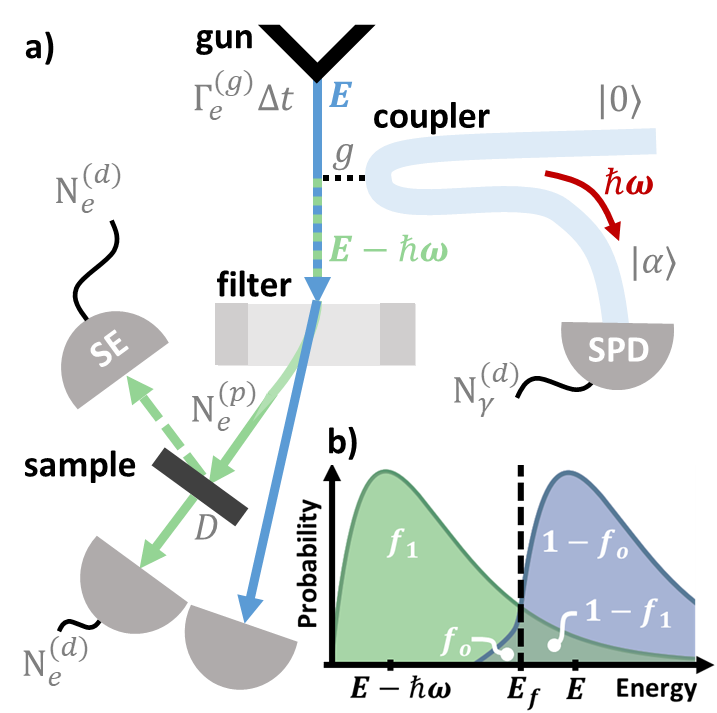}
    \vspace{-20pt}
    \caption{A heralded electron source. (a) Free electrons are generated by a standard electron source at a rate of $\Ge$ and then coupled to the evanescent field of a waveguide. Each electron produces, on average, $g\ll1$ photons. In one time bin (width $\Delta t$), $\Ng$ photons are detected with a single photon detector (SPD) and $\Np$ electrons are selected by an energy filter to be sent to the sample. Each of these electrons has probability $D$ of being counted by an electron detector (perhaps indirectly, via secondary electron bursts), resulting in $\Nd$ counts. (b) The interaction with the waveguide causes the electron beam to develop energy loss peaks spaced by the photon energy, $E_\gamma=\hbar\omega$. The energy filter removes the zero-loss component of the electron by rejecting energies $E>E_f$. The transmitted fractions of the zero- and single-loss peaks are $f_0$ and $f_1$, respectively.}
    \label{Fig:schematic}
\end{figure}

A schematic for a photon-heralded electron source is shown in Fig. \ref{Fig:schematic} (the supplementary material (\ref{app:symbol_table}) contains a table with a more complete set of symbols). Free electrons are generated from a standard electron gun, then focused and steered to pass within the evanescent field of a waveguide, generating photons as it passes. These photons are sent to a single photon detector, and the resulting electrons to an electron counting spectrometer.  The interaction creates an entangled electron-photon state. If the waveguide input is the ground state, then the output will have Poissonian photon-number statistics \cite{Kfir2019}.  Experimental demonstrations to-date have mean photon number (per electron) $g\ll 1$; however $g\gtrsim 1$ is possible in principle.  In this analysis, we assume that $g\ll 1$, consistent with current experimental capabilities.

In photon heralding, a typical setup involves a nonlinear medium which converts a higher energy pump photon to two lower energy ones: the signal and the idler. A photon detected in the idler channel `heralds' the presence of a photon in the signal channel \cite{Kwiat1995}. The traditional figure of merit for heralded single-photon sources is the Klyshko efficiency, which is the conditional probability of measuring a photon in the signal channel given the detection of a photon in the idler channel. The Klyshko efficiency for electron heralding can similarly be defined as the conditional probability of measuring an electron given the detection of a photon. However, this figure of merit does not penalize `false negative' events, where an electron arrives without a coincident photon. A low false negative rate is critical for applications like dose control. 

Instead, we will describe the heralding efficiency in terms of the fractional reduction of variance (FRV) in the estimate of the number of electrons $\Np$ that pass the energy filter, given the detection of $\Ng$ photons in a given time interval $[t,t+\Delta t]$. Assuming the source can be described with Poissonian statistics, which is generally valid for typical operating conditions \cite{Gover2012,Haindl2023, Meier2023}, 

\begin{equation}
    \FRV{\Np}{\Ng}=\kappa_{e}\kappa_{\gamma}
    \label{Eq:FRV}
\end{equation}

\noindent where $\kappa_{e}$ is the Klyshko efficiency for heralding electrons using photons and  $\kappa_{\gamma}$ is the Klyshko efficiency for heralding photons using electrons
\begin{equation}
    \kappa_{e}=\frac{g\eta_\gamma f_1}{g\eta_\gamma+\Ggdc/\Ge}, \;     \kappa_{\gamma}=\frac{g\eta_\gamma f_1}{gf_1+f_0}\ ,
\end{equation}
where $\eta_\gamma$ is the photon detection efficiency, $f_1$ is the probability that an electron belonging to the chosen sideband passes through the filter, $f_0$ is the probability that an electron outside of the chosen sideband passes through the filter, $\Ge=I/q$ is the average rate of electron production from the gun with current $I$ ($q$ is the electron charge magnitude), and $\Ggdc$ is the photon detector background count rate. A derivation of both equations is available in the section \ref{app:Source_Stat} in the supplement. 

To achieve a high FRV, both $\kappa_e$ and $\kappa_\gamma$ must be close to 1. This requires a highly efficient photon detector. While a strong electron-photon coupling is not necessarily required, it must be possible for the energy filter to efficiently isolate the chosen sideband, which has intensity proportional to $g$ and may be superposed with the tail of the zero-loss distribution (i.e. we need $f_1\sim 1$ and $gf_1 \gtrsim f_0$). In practice, this will require the sideband spacing (equivalently, the photon energy) to be much larger than the energy spread of the electron gun.

In the supplement we tabulate the FRV for various types of sources and detectors (Table \ref{Tab:params}) and calculate the maximum energy filter efficiency (section \ref{Sup:filter}). Based on that analysis, we will assume $\eta_\gamma=0.9$, $f_1=0.9$, and $f_0=0$. With $g=0.01$, $I<\SI{1}{\nano\ampere}$, and $\Ggdc<\SI{1}{\kilo\hertz}$, the effect of background counts is negligible. Then $\kappa_e\approx f_1$ and $\kappa_\gamma\approx\eta_\gamma$, simplifications we assume through the rest of this work.  



\textit{Dose Estimation} The task of dose calibration --- measuring the mean beam current beyond the beam forming apertures --- is difficult to do accurately in most electron-optical systems. Resolving the moment-to-moment Poissonian fluctuations, which ultimately requires non-destructive measurements of the beam, is currently impossible. In microscopy, dose estimation is important for in-situ electron microscopy and for controlling sample damage. In lithography, dose noise limits device yield and minimum feature size. 

Without heralding, some noise can be suppressed by sampling $\Nd$, the number of primary electrons registered in one time bin (perhaps indirectly, e.g. through secondary electrons bursts in SEM). If the probability of registering a primary electron is $D$, the FRV in the estimation of the dose is $\FRV{\Np}{\Nd}={D}/({1+\Gamma_{e}^{(dc)}/\Gamma_e^{(g)}fD})$ where $f \equiv gf_1+(1-g)f_0$ is the total probability that an electron from the gun passes the filter and $\Gedc$ is the electron detector dark count count rate.

When $D\approx 1$ and the dark count rate is negligible, $\FRV{\Np}{\Nd}= 1$. In other words, with a perfect electron detector and no loss in the sample, there is no shot noise and so no benefit to heralding. As derived in the supplement, the FRV for dose estimation obtained by combining data from both detectors is 
\begin{multline}
    \FRV{\Np}{\Nd,\Ng}=\\
    \kappa_\gamma+\frac{(1-\kappa_\gamma)D}{1+\Gedc/\Gebg}-k_\gamma(1-D)\frac{1-\kappa_e}{1-\kappa_eD}    
\end{multline}
where $\Gebg=\Ge (f-g\eta_\gamma f_1)D$ is the rate of additional background counts due to unheralded electrons. Using the photon detector, the FRV can be close to 1 even when the electron detection efficiency is low, which is the case for lithographic systems where the primary electron is used to expose a resist and secondary electrons (SEs) are not counted efficiently. When $D=0$, the FRV for dose estimation is identical to the FRV for heralding (FRV=$\kappa_{e}\kappa_{\gamma}$).



In electron lithography, Poissonian fluctuations in the beam current can cause errors in the written pattern. For a resist with critical areal dose density $\dose_c$ (charge per area required for full exposure), the expected relative dose error for a pixel of area $A$ is $1/\sqrt{\dose_c A/q}$. The error can be reduced by using larger pixels, which increases the minimum feature size, or using a less-sensitive resist \cite{yang2020real}. 
For a beam-current density $J$ and clock speed $C$, the maximum write speed is possible for $\dose_c\leq J/C$. For $\dose_c>J/C$, the writing speed becomes current-limited. As an example, suppose $J=\SI{100}{\ampere\per\cm\squared}$ and $C=\SI{100}{\mega\hertz}$. Then the maximum write speed is possible for resists with $\dose_c<\SI{1}{\micro\coulomb\per\cm\squared}$ or 6 electrons per $(\SI{10}{nm})^2$ pixel. To get less than 10\% dose error in each pixel, the resist sensitivity would need to be decreased by a factor of 16, slowing write speed proportionally. To achieve less than 10\% error using $(\SI{4}{nm})^2$ pixels, the write speed would need to be 100 times slower than the maximum clock speed. Often, it is not possible to choose a resist with the optimal critical dose due to issues of availability and process compatibility.

\begin{figure}[h]
    \centering
    \includegraphics[width=0.5\textwidth]{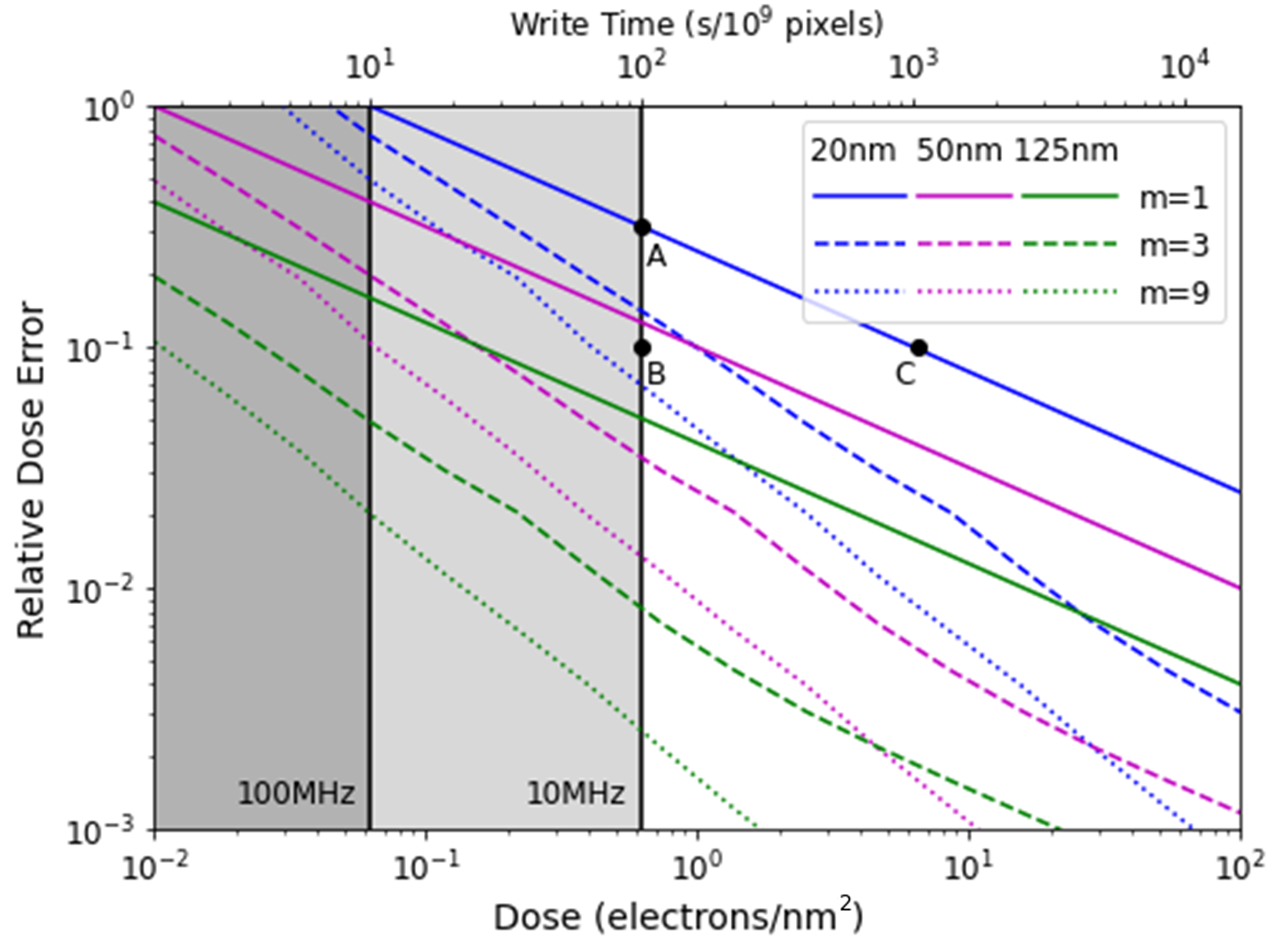}
    \vspace{-20pt}
    \caption{Relative dose error (standard deviation per mean dose) vs exposure dose for three different line widths (with 5 pixels per line). The top axis shows the current-limited write time using a \SI{100}{\ampere\per\cm\squared} electron source. In the grey regions, write time is limited by clock speed. The error can be  decreased by increasing line width or, at the cost of write time, by using less sensitive resist. Alternatively, dose error can be reduced using a heralded electron source and applying the dose into $m>1$ stages. Here, we assume the source is a FEG (so $f_1=1$) and the photon detector is an SNSPD (so $\kappa_\gamma=0.9$). The points $A$, $B$, and $C$ are referenced in the text.}

    \label{Fig:EL}
\end{figure}

Alternatively, the trade-off between dose error and speed can be controlled with heralding. To do so, the exposure is divided into $m$ stages and the dose applied at each stage is chosen based on the number of remaining stages and the estimated dose applied so far. This multi-pass, heralded electron lithography requires up to $m$ times as many clock cycles, slowing clock-limited write times. The source current is also reduced by a factor of $gf$, which slows current-limited write times. Despite this, in some circumstances heralding can significantly increase writing speed.

In Fig. \ref{Fig:EL}, we use the heuristic strategy of selecting the dose at each stage $k$ of the $m$ exposures according to $\dose_k=F(\dose_c-\sum_{k'<k}\hat{\dose}_{k'})$ where $\hat{\dose}_{k'}$ is the estimate of the dose applied at stage $k'$, and $F$ is a number between 0 and 1. The optimal choice for $F$ depends on $\dose_c$ and $m$. We calculate the (root mean square) dose error for $F=\{0.1,0.2,...0.9\}$ and save the best result. 

For a resist with critical dose of \SI{0.6}{electrons\per\nm\squared}, the dose error for a \SI{20}{\nm} linewidth (\SI{4}{\nm} pixels) is 30\% (see the point labeled $A$). Reducing the error below 10\% would require increasing the linewidth to more than \SI{60}{\nm} or finding a resist with one-tenth the sensitivity (see points $B$ and $C$, respectively). Alternatively, the dose could be controlled with a heralded source using a multi-pass exposure with $m=7$. If a small fraction of the pattern requires a linewidth of \SI{20}{\nm} while \SI{60}{\nm} is suitable for the rest, then electron lithography with a heralded source and sensitive resist is about $(6/2)^2=9$ times faster while maintaining relative error below 10\%.

\textit{Electron Heralding for Quantitative Microscopy} - Electron heralding has the potential to enhance electron microscopy when shot noise or dark counts are critical limitations. For example, when measuring sample transmissivity, the ability to distinguish loss events from fluctuations in the illumination can dramatically improve the signal to noise \cite{Agarwal2019}. This is possible with a heralded electron source, but some advantage is lost due to the reduced beam current. For a heralded source to be useful for faster acquisition, the signal to noise ratio (SNR) must increase sufficiently to compensate for the lost signal. Often, the SNR is limited by effects other than acquisition time and beam current, like detector well depth or sample charging and damage. If there is a critical dose above which the sample is unacceptably degraded, then an appropriate figure of merit is the information gain per electron. We will describe scenarios in SEM and bright field STEM where a heralded electron source can significantly improve the information gain per electron (or equivalently the SNR at constant dose).

A standard theoretical tool for optimizing measurements is the Fisher Information (FI), $\mathcal{I}(x)$, which is used in conjunction with the Cramer-Rao bound $\sigma_x^2\geq (\nd\mathcal{I}(x_0))^{-1}$ to determine the minimum measurement error $\sigma_x$ when estimating an unknown sample parameter $x$ (e.g. thickness, scattering mean free path, secondary electron (SE) yield, etc.) near a particular value $x=x_0 $\cite{ly2017tutorial, stoica1989music}. The quantity $\nd$ is a sample from the random variable $\Nd$. The advantage of this formalism is that we can determine the information value of a measurement without constructing an optimal method for estimating $x$ based on $\nd$. See section \ref{app:FischerInfo} of the supplement for a brief review of this topic.
We can re-write the Cramer-Rao bound in terms of the SNR = ${x}/{\sigma_x}\leq x\sqrt{\nd\mathcal{I}(x_0)}$.

The fractional increase in Fisher information achieved by heralding is
\begin{equation}
    \frac{\mathcal{I}(x_0|\Nd,\Ng)}{\mathcal{I}(x_0|\Nd)}=1+\kappa_\gamma\left(\frac{1+\Gedc/\Gebg}{1-D(x_0)\kappa_e}-1\right) \text{,}
    \label{Eq:FGImain}
\end{equation}
 By examining equation \ref{Eq:FGImain}, we can see two potential limiting cases where the fractional increase in information is large: when dark count rate is large ($\Gedc/\Gebg\gg1$) or when the contrast is low ($D(x_0)\sim 1$). In the remainder of this letter, we examine specific scenarios where heralding may be beneficial for electron microscopy.


\begin{figure}[t]
    \centering
    \includegraphics[width=0.45\textwidth]{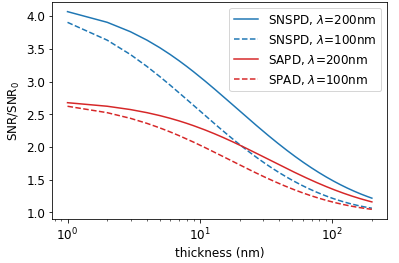}
    \vspace{-10pt}
    \caption{Fractional increase in bright-field STEM SNR vs sample thickness with heralding at constant dose for different photon detectors and electron mean free paths ($\lambda$).}
    \label{Fig:STEM}
\end{figure}

\textit{High Energy Electron Microscopy} - 
In bright field (BF) STEM, an aperture at the backfocal plane of the objective lens removes electrons which scatter to large angles as they pass through the sample. One possible motivation for such a measurement is to estimate the thickness $t$ of a sample with a known, material-dependent scattering mean-free-path $\lambda$. If the electron detector has quantum efficiency $\eta_e$, then the detection probability is $D(t)=\eta_e (1-e^{-\lambda/t})$.
In Fig. \ref{Fig:STEM}, we show the fractional increase in SNR (the square root of Eq. \ref{Eq:FGImain}) relative to the unheralded case as a function of $t$ for $\eta_e=0.95$.
With an SNSPD,
heralding doubles the SNR ratio at constant dose for samples with $\lambda=\SI{100}{\nm}(\SI{200}{\nm})$ and $t<\SI{20}{\nm}(\SI{42}{\nm})$. The information added by the heralding system in this case is equivalent to the information collected by a high-efficiency annular dark field detector covering all scattering angles excluded by the bright field detector.  



For other high-energy electron-microscopy modalities, it is more difficult to use a heralded source advantageously. Transmission electron microscopy and EELS use pixelated detectors with readout speeds far too low to preserve correlations with the photon detector signal, but heralding could improve dose estimation as discussed in the previous section for in-situ experiments. Heralding could also help reduce noise backgrounds for EDS and CL to the extent that the noise is uncorrelated with the primary electrons. 

\begin{figure*}[t]
    \centering
    \includegraphics[width=0.99\textwidth]{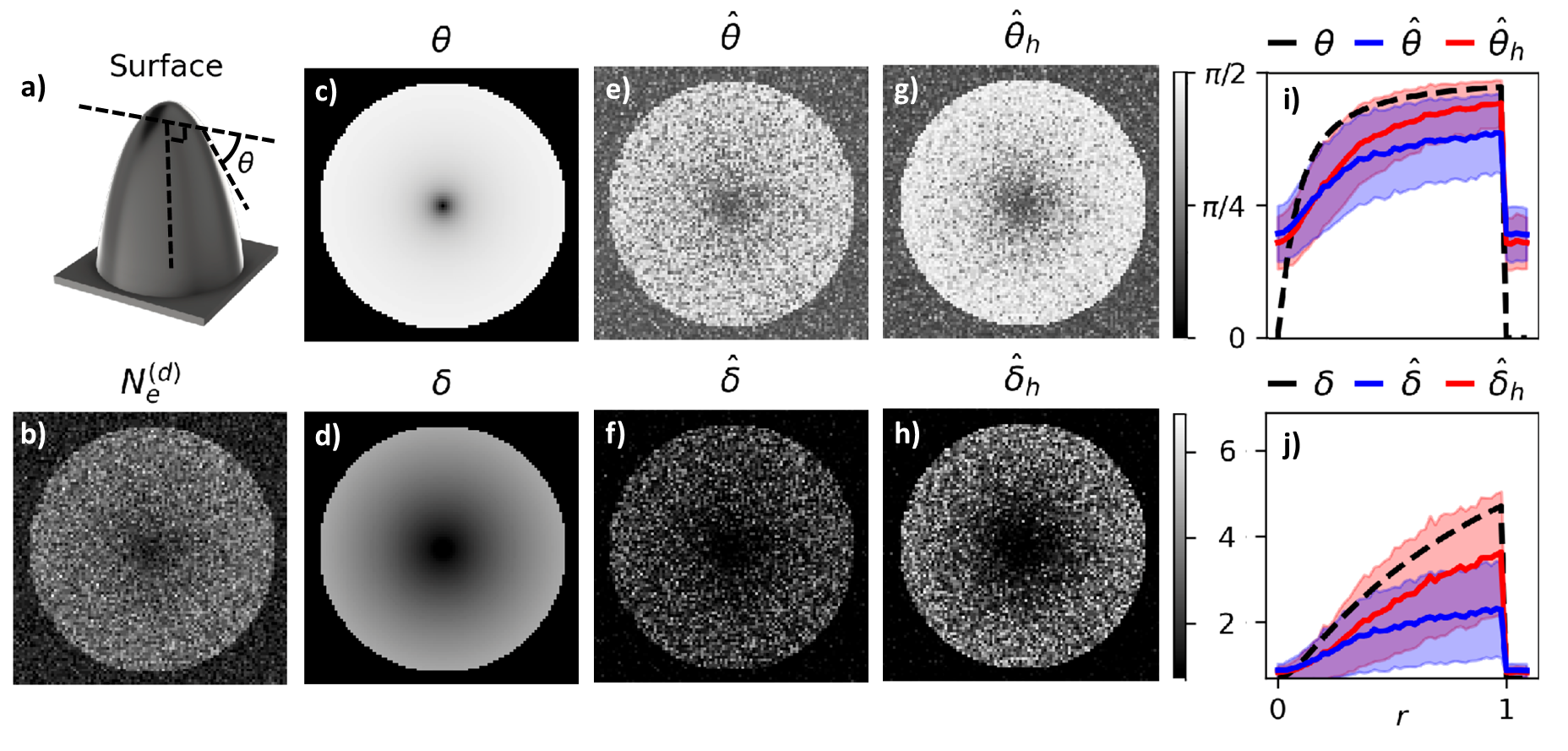}
    \vspace{-20pt}
    \caption{Simulated SEM image showing the effects of heralding. (a) Shows a rendering of the sample, which choose to be a parabola \SI{1.35}{\um} tall with a \SI{450}{\nm} base. (b) Simulated SEM micrograph of the sample when an average dose of 10 electrons is applied to each $(\SI{10}{\nm})^2$ pixel. (c-d) Ground truth of the profile angle ($\theta$) and expected SE yield ($\delta$), respectively. (e-f) show estimates for $\theta$ and $\delta$ without heralding. (g-h) show the improved estimate with heralding. (i-j) show the values of $\theta$, $\delta$, and the expected values for their estimators (calculated from 1000 samples) for the radial profile of the the feature. Shaded areas cover one standard deviation above and below each mean.  }
    \label{Fig:SEM_sim}
\end{figure*}

\textit{Quantitative SEM} - The interpretation of SEM images can change dramatically based on the beam energy and sample composition and morphology. For example, more secondary electrons escape from the interaction volume near the edges of nanostructures, causing a bright halo and a reduction in contrast. This so-called edge effect \cite{Wells1978,Matsukawa1974} can be partially ameliorated using a heralded electron source. To show this we model the tilt-dependent SE yield as $\delta(\theta)={\delta_0\delta_1}/{[\cos(\theta)(\delta_1-\delta_0)+\delta_0]}$ where $\theta$ is the surface tilt, $\delta_0$ is the SE yield for a horizontal surface ($\theta=0$), and $\delta_1$ is the SE yield at the edge of a tall vertical step \cite{Postek2014}. We assume the detector clicks at most once for each primary electron, which is true when a single SE saturates the detector for longer than the spread in SE arrival times, and so counting secondary electrons is not possible\cite{Agarwal2023}. We also assume that the SE yield is Poissonian (with mean $\delta$). Then we write the probability of detection as $D(\delta)=1-e^{-\eta_e \delta}$.

When the background count rate is small, the increase in information from heralding is proportional to $e^{\eta_e \delta}$. Heralding is especially advantageous when the SE yield is much larger than 1 (i.e. where $\theta\sim \pi/2$), so contrast is low due to detector saturation. In such cases, the most information-rich events are primary electrons which fail to produce a SE. Without heralding, these events cannot be recognized. 

Figure \ref{Fig:SEM_sim} shows a simulation of parameter estimation in an SEM. It neglects some aspects of image formation in SEM (e.g. shadowing and point spread function), but it provides an approximate visual comparison of quantitative SEM with and without heralding. At each pixel, the number of incident primary electrons is drawn from a Poisson distribution with mean 10. Then numbers are drawn from a multinomial distribution to determine the number of coincident and non-coincident events at the electron and photon detectors. Without heralding the estimates of the surface tilt and SE yield are $\hat{\theta}$ and $\hat{\delta}$. With heralding the estimates are $\hat{\theta}_h$ and $\hat{\delta}_h$. See section \ref{Sec:SEM_Est} of the appendix for more details. 

Prior probability distributions for $\theta$ and $\delta$ must be specified. We assume a uniform distribution for $\theta$ (which induces a non-uniform distribution for $\delta$). As the true surface tilt is not well-represented by a uniform distribution, these estimators are biased (e.g. $|\langle\hat{\delta}-\delta\rangle|>0$ at finite dose). As seen in the radial plots on the right of the figure, heralding reduces this bias (an effect not captured by the Fisher information analysis). 

Heralding reduces the error for the estimation of $\theta$ and $\delta$ by a factor of 1.6 and 1.3, respectively. Equivalently, the dose required to reach a prescribed estimate error for $\delta$ and $\theta$ is reduced by a factor of 2.6 and 1.8 respectively. Compositional analysis could also be improved by heralding, but the advantage would only be exist when SE yields are high enough to saturate the detector. 

\textit{Conclusion and outlook} - We have described a realistic heralded electron source, calculated its statistics, and estimated its effectiveness in specific applications of STEM, SEM, and electron lithography. A general figure of merit for describing the effectiveness of a heralded source is the FRV, which is proportional to the Klyshko efficiencies for both electron and photon heralding. To achieve a high FRV in practice will not necessarily require additional advances in sources, detectors, energy filters, or electron-photon coupling structures, but it will require sophisticated engineering to integrate state-of-the-art implementations of each of these systems.

The goal of this analysis was to identify realistic circumstances where electron beam technologies may be improved by a heralded source. We chose to examine a series of minimal models which may motivate more detailed investigations of each potential application in the future. 

Multi-pass heralded electron lithography would have reduced beam current and may require more clock cycles. However, it enables dynamic control of the trade-off between speed and noise. Using a single layer of resist, heralded electron lithography could quickly expose regions of low detail, then apply a low-noise multi-stage exposure to areas where noise could limit device yield. 

Not all forms of microscopy benefit from heralding. For example, in phase contrast electron microscopy, information is extracted from the relative brightness of electron detector pixels. However, for amplitude (i.e. bright field) microscopy, heralding makes it possible to detect events where primary electrons fail to arrive at the electron detector. These events are particularly informative when they are rare (i.e. low-contrast imaging). For BF STEM of thin samples, a heralded electron source can more-than double the SNR at constant dose. Similarly, heralding is beneficial in SEM for low-contrast imaging conditions. An electron microscope equipped with a heralding system would likely be operated in a high-current, unheralded mode for alignment, focusing, and feature-finding, then switched to a low-current, heralded mode for dose calibration and enhanced performance.

This work clarifies the specific conditions in which improvement from heralding can be expected. It also motivates more detailed analysis and design of heralded electron sources, which have the potential to become indispensable to next-generation electron-optical systems.


\begin{acknowledgments}
We are grateful to Dr. Felix Ritzkowsky, Owen Medeiros, and Camron Blackburn for helpful discussion and input on the manuscript.  This material is based upon work supported by the National Science Foundation under Grant No. 2110535 (MIT) and No. 2110556 (UC Davis).  
\end{acknowledgments}

\bibliography{refs} 

\newpage
\appendix
\appendix

\section{Table of Symbols}

\label{app:symbol_table}


\begin{tabular}{cp{.35\textwidth}l}
\textbf{symbol} & \textbf{meaning} \\
\hline
$\Delta t$ & size of one time bin\\
$\Neg$ & random variable (RV) describing the number of electrons produced by the gun in one time bin (samples from random variables are written in lower case)\\
$\Np$ & RV describing the number of electrons which pass the filter in one time bin\\
$\Nd$ & RV describing the number of primary electrons which are registered by the electron detector (perhaps by generating one or more secondary electrons) in one time bin\\
$\Ng$ & RV describing the number of photons which are registered by the single photon detector in one time bin\\
$\Ge$ & rate at which electrons are produced from gun (after apertures)\\
$\Gedc$ & electron detector dark count rate\\
$\Gebg$ & electron detector background count rate from unheralded electrons\\
$\Ggdc$ & photon detector background count rate\\
$g$ & expected number of photons produced per electron\\
$f_0$ & probability that an electron outside of the chosen sideband passes through the filter\\
$f_1$ & probability that an electron belonging to the chosen sideband passes through the filter\\ 
$f$ & total probability that an electron passes through the filter\\
$D$ & probably the electron detector (secondary or primary) registers a count given that a primary electron passed the filter\\
$\kappa_e$ & Klyshko efficiency for heralding electrons using photons\\
$\kappa_\gamma$ & Klyshko efficiency for heralding photons using electrons\\
$\mathcal{I}$ & Fisher information\\
FRV & fractional reduction in variance
\end{tabular}

\section{Statistics of a heralded electron source}\label{app:Source_Stat}
In this section, we calculate the fractional reduction in variance achievable with the heralded source described in the main text. In general, the superscripts $(g)$, $(p)$, and $(d)$ will distinguish between quantities relating to the electron gun, the exit of the energy filter, and the electron/photon detectors, respectively. 

To begin, we assume the photon detector is capable of resolving arrival times into bins of size $\Delta t$ with $\Delta t\Ge=\epsilon\ll 1$, where $\Ge=I/q$ is the average rate of electron production, $I$ is the beam current, and $q$ is the electron charge. With this assumption, each of the events in the system can be described as Bernoulli (binary) random variables. Let $\Neg$ be a random variable describing the number of electrons produced by the gun in the time interval $[t,t+\Delta t]$. While space charge and Fermionic particle statistics can in principle result in an anti-bunched electron beam, this effect is typically very small and has only recently been observed \cite{Gover2012,Haindl2023, Meier2023}. Therefore we will assume electron arrival times are uncorrelated so $\Neg$ is Poissonian with mean and variance $\mean{\Neg}=\Delta t\Ge$. Let $\Np$ be a random variable describing the number of electrons which pass through the energy filter of the heralding system. Without access to data from the single photon detector $\Np$ is Poissonian. Let $\Ng$ be a random variable describing the number of detected photons (it too is Poissonian with mean and variance $g\Delta t\Ge$). Then 
\begin{align}
    p(\Neg=1)&\equiv p(e_g)=\Ge\Delta t+\mathcal{O}(\epsilon^2)\nonumber\\
    p(\Neg=0)&\equiv p(\neg e_g)=1-\Ge\Delta t\nonumber
\end{align}
and
\begin{align}
    p(\Ng=1)&\equiv p(\gamma_d)=g\eta_\gamma\Ge\Delta t+\Ggdc\Delta t\nonumber\\
    &\qquad-(\Ggdc/\Ge)\mathcal{O}(\epsilon^2)\nonumber\\
    p(\Ng=0)&\equiv p(\neg \gamma_d)=1-p(\gamma_d)\nonumber
\end{align}
where $g$ is the average number of photons produced per electron (we will assume $g\ll 1$), $\eta_\gamma$ is the photon system detection efficiency, and $\Ggdc$ is the photon detector background count rate. 
The term proportional to $\epsilon^2$ comes from simultaneous signal and background events and is small as long as $\Ggdc\ll\Ge$. We will assume this is true. Based on the definitons of $g$ and $\eta_\gamma$, we can also write the conditional probability $p(\gamma_d|e_g)=g\eta_\gamma$.

After interacting with the coupler, the electron beam is energy-filtered. We will assume the filter uses an energy-selecting slit which passes electrons with energy $E_0<E<E_1$. If $\mathcal{S}(E)$ is the energy distribution of the electron beam at the source and $E_\gamma$ is the energy of a signal photon (we will assume the energy distribution of the photons is very narrow compared to $\mathcal{S}(E)$), then let
\begin{align}
    f_0&=\int_{E_0}^{E_1} dE\mathcal{S}(E)\nonumber\\
    f_1&=\int_{E_0}^{E_1} dE\mathcal{S}(E-E_\gamma)\nonumber
\end{align}
We can interpret $f_0$ as the probability that an electron passes the filter given that it didn't produce a photon and $f_1$ as the probability that an electron passes the filter given that it did lose energy $E_\gamma$ to a photon. Then
\begin{equation}
    f\equiv p(e_p|e_g)=gf_1+(1-g)f_0\nonumber
\end{equation}
is the total probability that an electron which emerges from the gun passes the filter.

In any given time bin, there are four possible outcomes
\begin{align}
    & \Np=1,\ \Ng=1\quad\text{true positive} \nonumber\\
    & \Np=0,\ \Ng=1\quad\text{false positive} \nonumber\\
    & \Np=1,\ \Ng=0\quad\text{false negative} \nonumber\\
    & \Np=0,\ \Ng=0\quad\text{true negative} \nonumber
\end{align}
with probabilities
\begin{align}
    p(e_p|\gamma_d)\quad\text{true positive}\qquad & p(\neg e_p|\gamma_d)\quad\text{false positive} \nonumber\\
    p(e_p|\neg \gamma_d)\quad\text{false negative} \qquad & p(\neg e_p|\neg \gamma_d)\quad\text{true negative} \nonumber
\end{align}
We can also condense these outputs into a single figure of merit in the form of the conditional variance
\begin{align}
    \Var{\Np|\Ng}&\equiv\sum_n p(\Ng=n) \Var{\Np|\Ng=n}\nonumber\\
   =&p(\gamma_d)\Var{\Np|\gamma_d}\nonumber\\&+p(\neg\gamma_d)\Var{\Np|\neg\gamma_d}\nonumber
\end{align}
with
\begin{align}
    \Var{\Np|\gamma_d}&=p( e_p|\gamma_d)\bigg(1-p(e_p|\gamma_d)\bigg)\nonumber\\
    \Var{\Np|\neg\gamma_d}&=p( e_p|\neg\gamma_d)\bigg(1-p( e_p|\neg\gamma_d)\bigg)\nonumber\ .
\end{align}
In order to calculate these conditional probabilities, we use the above relations to obtain
\begin{align}
    p( e_p|\gamma_d)&=\frac{p(\gamma_d,e_p)}{p(\gamma_d)}=\frac{\Ge \Delta tg\eta_\gamma f_1}{g\eta_\gamma\Ge\Delta t+\Ggdc\Delta t}\nonumber\\
    p( e_p|\neg\gamma_d)&=\frac{p(\neg\gamma_d,e_p)}{p(\neg\gamma_d)}=\frac{\Ge\Delta t (f-g\eta_\gamma f_1)}{1-g\eta_\gamma\Ge\Delta t-\Ggdc\Delta t}\nonumber
\end{align}
giving
\begin{align}
    \frac{\Var{\Np|\Ng}}{\Delta t\Ge f}&=1-\frac{\left(gf_1\eta_\gamma\Ge\right)^2}{f\Ge\left(g\eta_\gamma\Ge+\Ggdc\right)}+\mathcal{O}(\epsilon^2)\ .\nonumber
\end{align}

The intrinsic variance of the source (the variance without information from the photon detector) is
\begin{equation}
    \Var{\Np}=\Delta t\Gamma_g^{(e)}f+\mathcal{O}(\epsilon^2)\ .\nonumber
\end{equation}
Incorporating the information from the photon detector, the fractional reduction in variance (FRV) is, to first order in $\epsilon$,
\begin{align}
    \FRV{\Np}{\Ng}&\equiv\frac{\Var{\Np}-\Var{\Np|\Ng}}{\Var{\Np}}\nonumber\\
    &=\frac{\left(gf_1\eta_\gamma\Ge\right)^2}{\left(f\Ge\right)\left(g\eta_\gamma\Ge+\Ggdc\right)}\nonumber\\
    &=\frac{\Gamma^{(e\gamma)}}{\Gamma^{(e)}}\frac{\Gamma^{(e\gamma)}}{\Gamma^{(\gamma)}}\nonumber\\
    &=\kappa_\gamma\kappa_e
\end{align}
where $\Gamma^{(e\gamma)}$, $\Gamma^{(e)}$, and $\Gamma^{(\gamma)}$ are the rates of coincident events, electron detector events, and photon detector events, respectively. 

Without heralding, the FRV is 0. With perfect heralding, the FRV is 1. In terms of the system parameters described above,

\begin{equation}
    \kappa_{e}=\frac{gf_1\eta_\gamma\Ge}{g\eta_\gamma\Ge+\Ggdc} \;,\;     \kappa_{\gamma}=\frac{g\eta_\gamma f_1}{f}\ .
\end{equation}

where $\Ggdc$ and $\eta_\gamma$ are the photon detector background count rate and detection efficiency, respectively.
\begin{table}
\label{Tab:params}
\centering
\begin{tabular}{|c|c|c|c|c|c|c|}
    \hline
    system  & $\eta_\gamma$ & $f_0$ & $f_1$ & $\kappa_e$ &  $\kappa_{\gamma}$ & FRV$(\Np|\Ng)$\\
    \hline
    SNSPD+FEG  & 0.9  & 0 & 1 & 1 & 0.9 & 0.9\\
    \hline
    SNSPD+SG & 0.9  & 0 & 0.9 & 0.9 & 0.9 & 0.8\\
    \hline
    SPAD+FEG & 0.7  & 0 & 1 & 1 & 0.7 & 0.7\\
    \hline
    SPAD+SG & 0.7  & 0 & 0.9 & 0.8 & 0.7 & 0.6\\
    \hline
\end{tabular}
\caption{Dimensionless parameters describing potential heralding systems using either a field emission gun (FEG) with $\Delta E=\SI{0.3}{\electronvolt}$ or a schottky gun (SG) with $\Delta E=\SI{0.7}{\electronvolt}$; and using either a Superconducting Nanowire Single Photon Detector (SNSPD) with $\eta_\gamma=0.9$ or a Single Photon Avalanch Detector (SPAD) with with $\eta_\gamma=0.7$. In all cases, we assume the electron-photon coupling is $c=0.01$, the beam current is $I=\SI{1}{\nano\ampere}$, the photon detector background count rate is $\Ggdc=\SI{1}{\kilo\hertz}$, and the photon energy is $E_\gamma=\SI{1.1}{\electronvolt}$.   }
\end{table}

\subsection{Optimal filter parameters}
\label{Sup:filter}
Here we justify $f_1=1$ for a field emission gun and $f_1=0.9$ for a Schottky emitter. First we assume the tip current is low enough such that the energy distribution of the emitted electrons is not distorted by space charge effects. Adapting this from Riemer \cite{reimer2013transmission} we have

\begin{equation}
N(E) \mathrm{d} E=\frac{E}{\left(k T_{\mathrm{c}}\right)^2} \exp \left(-E / k T_{\mathrm{c}}\right) \mathrm{d} E\nonumber
\end{equation}

By setting the tip energy spread to 0.7 eV (for a Schottky emitter, which corresponds to a \SI{3320}{\kelvin} tip temperature), we find the overlap of a single \SI{1.1}{\electronvolt} photon sideband with the zero-loss distribution is 0.1, as shown in Fig. \ref{fig:energyThresh} below. An otherwise ideal energy filter could then have $f_1 = 0.9$.

\begin{figure*}[t]
    \centering
    \includegraphics[width=\textwidth]{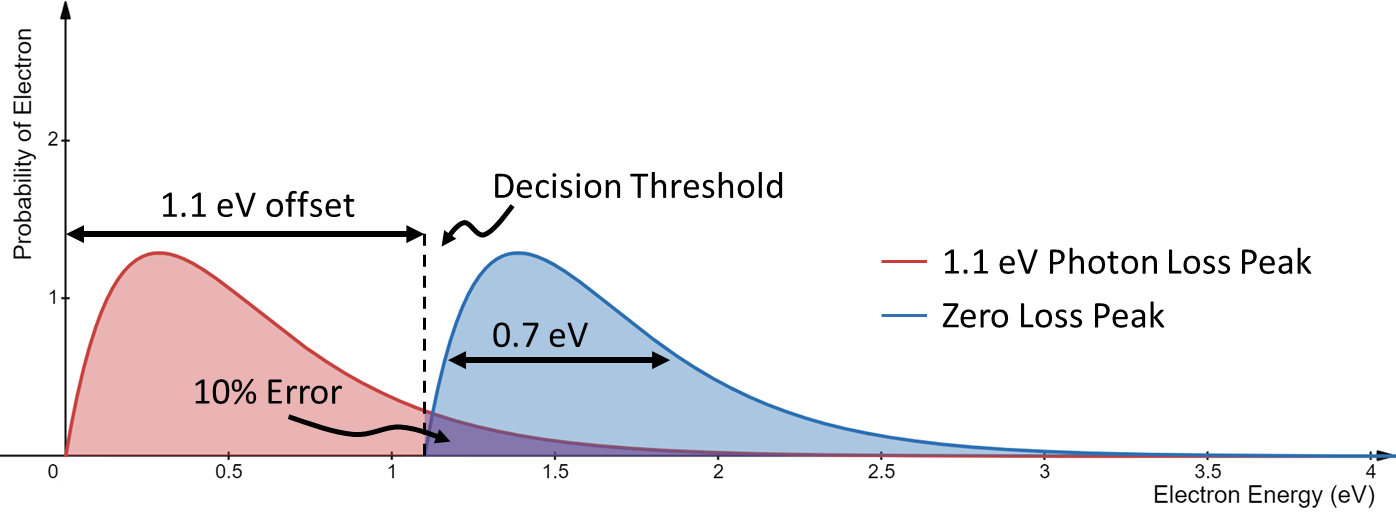}
    \caption{Schematic of the electron threshold. If we have an 0.7~eV spread with 1.1~eV separation, then there is a 10\% chance that an electron that lost a photon will be identified as one that did not. This is found by integrating the curve above the decision threshold line, shown in purple.}
    \label{fig:energyThresh}
\end{figure*}

For the field emitter, the energy spread is so small that this integral is practically zero, and so $f_1=1$.

\section{Dose Estimation}\label{app:Dose_Est}
Here we calculate the reduction in the variance of $\Np$ achievable using both a photon detector and an electron detector. We will assume the electron detectors are fast enough to count primary electrons. This assumption may not be appropriate for the pixelated detectors used, for example, in transmission electron microscopy, but it is possible for STEM\cite{Peters2023} and SEM\cite{Agarwal2023} detectors with careful calibration.

First, to find the reduction in variance using just the electron detector, we first need
\begin{align*}
     \Var{\Np\middle|\Nd}=&\sum_{\Nd}p(\Np)\Var{\Np\middle|\Nd}\\
    =&p(e_d)\left[p(e_p|e_d)\big(1-p(e_p|e_d)\big)\right]\\
    &+p(\neg e_d)\left[p(e_p|\neg e_d)\big(1-p(e_p|\neg e_d)\big)\right]\\
    =&p(e_d)\left[p(e_p|e_d)\big(1-p(e_p|e_d)\big)\right]\\
    &+p(e_p|\neg e_d)+\mathcal{O}(\Delta t^2)
\end{align*}
And since
\begin{align*}
    p(e_p|e_d)&=\left[\frac{p(e_p)}{p(e_d)}p(e_d|e_p)\right]\\
    &=\frac{\Ge f}{\Ge fD+\Gedc}D+\mathcal{O}(\Delta t)\\
    &=\frac{1}{1+r_e} 
\end{align*}
where
\begin{equation}
    r_e=\frac{\Gedc}{\Ge fD}
\end{equation}
and
\begin{align*}
    p(e_p|\neg e_d)&=\left[\frac{p(e_p)}{p(\neg e_d)}p(\neg e_d|e_p)\right]\\
    &=\Delta t\Gamma_g^{(e)}f(1-D)+\mathcal{O}(\Delta t^2)
\end{align*}
therefore 
\begin{equation*}
    \Var{\Np|\Nd}=\Delta t\Gamma_g^{(e)}f\left(1-\frac{D}{1+r_e}\right)+\mathcal{O}(\Delta t^2)
\end{equation*}
At this point, we can find the fractional reduction of $\Np$ using $\Nd$:
\begin{equation*}
    \text{FRV}(\Np|\Nd)=1-\frac{\Var{\Np|\Nd}}{\Var{\Np}}=\frac{D}{1+r_e}
\end{equation*}

Now we will calculate the additional variance reduction achievable with a heralding system. First,
\begin{align*}
&\Var{\Np\middle|\Nd,\Ng}\\
&=\sum_{\nd,\ng}p(\nd,\ng)\Var{\Np\middle|\nd,\ng}\\
&=p(e_d,\gamma_d)\left(p(e_p|e_d,\gamma_d)(1-p(e_p|e_d,\gamma_d))\right)\nonumber\\
&+p(e_d,\neg\gamma_d)\left(p(e_p|e_d,\neg\gamma_d)(1-p(e_p|e_d,\neg\gamma_d))\right)\nonumber\\
&+p(\neg e_d,\gamma_d)\left(p(e_p|\neg e_d,\gamma_d)(1-p(e_p|\neg e_d,\gamma_d))\right)\nonumber\\
&+p(\neg e_d,\neg \gamma_d)\left(p(e_p|\neg e_d,\neg\gamma_d)(1-p(e_p|\neg e_d,\neg\gamma_d))\right)
\end{align*}
Since
\begin{equation*}
    p(e_p|e_d,\gamma_d)=1+\mathcal{O}(\Delta t^2)
\end{equation*}
 and\begin{equation*}
    \mathcal{O}(1-p(\neg e_d,\neg\gamma_d))=\mathcal{O}(p(e_p|\neg e_d,\neg \gamma_d))=\mathcal{O}(\Delta t)
\end{equation*}
we can further simplify
\begin{align*}
&\Var{\Np\middle|\Nd,\Ng}\\
&=p(e_d,\neg\gamma_d)\left(p(e_p|e_d,\neg\gamma_d)(1-p(e_p|e_d,\neg\gamma_d))\right)\nonumber\\
&+p(\neg e_d,\gamma_d)\left(p(e_p|\neg e_d,\gamma_d)(1-p(e_p|\neg e_d,\gamma_d))\right)\nonumber\\
&+p(e_p|\neg e_d,\neg\gamma_d)+\mathcal{O}(\Delta t^2)
\end{align*}
so we'll need to find five probabilities: $p(e_d,\neg \gamma_d)$, $p(\neg e_d,\gamma_d)$, and three permutations of $p(e_p|(\neg)e_d,(\neg)\gamma_d)$.

Let's start with $p(e_d,\neg \gamma_d)$. The expression should contain one term representing a dark count at the electron detector and another where a real electron strikes the detector but either didn't create a photon or the photon was lost.
\begin{equation*}
    p(e_d,\neg \gamma_d)=\Delta t\Gedc+\Delta t\Ge\left(f-gf_1\eta_\gamma\right)D+\mathcal{O}(\Delta t^2)
\end{equation*}

Using similar reasoning, 
\begin{equation*}
    p(\neg e_d, \gamma_d)=\Delta t\Ggdc+\Delta t\Ge g\eta_\gamma\left(1-f_1D\right)+\mathcal{O}(\Delta t^2)
\end{equation*}

Now the conditional probabilities. First, we have
\begin{equation*}
    p(e_p|\neg e_d,\neg\gamma_d)=\Delta t\Ge(f-gf_1\eta_\gamma)(1-D)+\mathcal{O}(\Delta t^2)   
\end{equation*}
Then,
\begin{align*}
    p(e_p|\neg e_d,\gamma_d)&=\frac{p(e_p,\neg e_d,\gamma_d)}{p(\neg e_d,\gamma_d)}\\
    &=\frac{\Ge g\eta_\gamma f_1\left(1-D\right)}{\Gamma_{dc}^{(\gamma)}+\Gamma_g^{(e)}g\eta_\gamma\left(1-f_1D\right)}
\end{align*}
Finally,
\begin{align*}
    p(e_p|e_d,\neg\gamma_d)&=\frac{p(e_p,e_d,\neg\gamma_d)}{p(e_d,\neg\gamma_d)}\\
    &=\frac{\Gamma_g^{(e)}\left(f-gf_1\eta_\gamma\right)D}{\Gamma_{dc}^{(e)}+\Gamma_g^{(e)}\left(f-gf_1\eta_\gamma\right)D}\\
    &=\frac{f-gf_1\eta_\gamma}{fr_e+f-gf_1\eta_\gamma}
\end{align*}
Putting everything together, we have

\begin{align}
    \frac{\Var{\Np\middle|\Nd,\Ng}}{\Delta t\Ge f}&=(1-\kappa_\gamma)(1-D)\nonumber\\
    &+\frac{(1-\kappa_\gamma)Dr_e}{1+r_e-\kappa_\gamma}\nonumber\\
    &+k_\gamma(1-D)\frac{1-\kappa_e}{1-\kappa_eD}
\end{align}

\subsection{Dose Estimation for Electron Lithography}\label{app:Lith_Est}

As described in the main text, a multi-pass procedure can be used to implement dose control. In our, strategy the dose applied at stage $k$ is
\begin{equation}
    \dose_k=F(\dose_c-\sum_{k'<k}\hat{\dose}_{k'})
\end{equation}
where $d_c$ is the target dose and $\hat{d}_k$ is the estimated dose at stage $k$. Without any information from the photon detector, we use $\hat{d}_k=n_{\gamma,k}/\kappa_\gamma$ where $n_\gamma$ is the number of photons detected at stage $k$. For the purpose of simulation, we draw $n_\gamma$ from a binomial distribution $\mathcal{B}(n_e;\kappa_e)$ where $n_e$ is the number of electrons actually delivered to the sample, which itself is a random number drawn from a Poisson distribution with mean $d_k$. The optimal value of constant $F$ depends on the total number of dose stages $m$ and on $d_c$. 

\section{Review of Relevant Concepts related to Fisher Information}\label{app:FischerInfo}

The goal of a quantitative measurement is to estimate an unknown sample parameter, traditionally labeled $\theta$. In STEM, $\theta$ could represent sample thickness or scattering cross-section. In SEM, $\theta$ could represent the secondary electron yield or surface tilt. A function $\hat{\theta}(X)$ which uses measurement data $X$ to produce an estimate of $\theta$ is called an estimator. We consider $X$ to be a random variable and the variance $\Var{\hat{\theta}}=\mean{(\hat{\theta}-\theta)^2}$ is the square of the measurement error.

It is possible to place an upper bound on the measurement error without formulating $\hat{\theta}$ using the Cramer-Rao bound
\begin{equation}
    \Var{\hat{\theta}}\leq \left(N\mathcal{I}(\theta_0)\right)^{-1}
\end{equation}
where
\begin{equation}
    \mathcal{I}(\theta_0)=\mathbb{E}\left\{\left(\frac{\partial}{\partial \theta} 
    \log X(\theta)\right)^2\middle|\theta_0\right\}
\end{equation}
is the Fisher information (FI). The Cramer-Rao bound applies only to unbiased estimator (with zero expected error). 

As an example, a Bernoulli random variable with probability mass function $p(k;\theta)=\theta^k(1-\theta)^{1-k}$ has FI
\begin{equation}
    \mathcal{I}_B=\frac{N}{\theta(1-\theta)}
\end{equation}
This means that after $N$ samples of the distribution, the error of the best unbiased estimator is $\sqrt{N\theta(1-\theta)}$. Similarly, for a random variable with a Poisson distribution $p(k;\theta)=e^{-\theta}\theta^k/k!$,
\begin{equation}
    \mathcal{I}_P=\frac{N}{\theta}
\end{equation}
so after $N$ samples of the distribution, the error of the best unbiased estimator is $\sqrt{N\theta}$. Notice that when estimating the parameter of a Bernoulli distribution, the FI diverges for large and small values of $\theta$. But when estimating the parameter of a Poisson distribution, the FI diverges only when $\theta=0$. 

\section{Fisher Information for Electron Microscopy}\label{app:ParamEst}
We now to estimate an unknown parameter $x$, which determines $D(x)$, the probability that an electron produced by the heralded source triggers the electron detector. The Fisher Information (FI), $\mathcal{I}(x)$, added per time bin is

\begin{align*}
    &\mathcal{I}(x_0|\Nd,\Ng)=\\
    &\sum_{\nd,\ng=0}^1\frac{1}{p(\nd,\ng)}\left(\partial_{x}p(\nd,\ng)\middle|_{x=x_0}\right)^2
\end{align*}
As above, we simplify the calculation of $\mathcal{I}$ by assuming that the electron detector operates in a counting regime (it has a binary response to each primary electron) with $D(x)$ the probability that an electron which is produced by the heralded source is counted by the electron detector. Without heralding, the FI for values of $x$ near $x_0$ gained from sampling only $\Nd$ (i.e. without heralding) is 
\begin{equation*}
    \mathcal{I}(x_0|\Nd)=\frac{\Delta t\Ge\left(\partial_x D(x)\middle|_{x_0}\right)^2}{D(x)(1+\Gedc/\Gebg)}
\end{equation*}
where $\Gebg$ is the rate at which unheralded electrons arrive at the detector. Using data from the photon detector, we have

\begin{equation}
    \mathcal{I}(x_0|\Nd,\Ng)=\Delta t\Ge\left(\partial_x D(x)\middle|_{x_0}\right)^2\left(I_\gamma+ I_e\right), 
\end{equation} \label{eq:FI_h}
 where 
\begin{equation}
    I_\gamma = \frac{\kappa_\gamma}{D(x_0)(1-D(x_0)\kappa_e)}
\end{equation}

and
\begin{equation}
    I_e = \frac{1-\kappa_\gamma}{D(x_0)(1+\Gedc/\Gebg)}.
\end{equation}

The first term in Eq. \ref{eq:FI_h} represents the information associated with photon detections (both with and without a coincident electron detection) and is similar in form to the FI associated with a binomial random variable with success probability $D(x)$. The second term represents the information associated with electron detections which were not heralded and is similar in form to the FI associated with a Poisson random variable. Combining these two expressions above, the fractional gain in information is
\begin{equation}
    \frac{\mathcal{I}(x_0|\Nd,\Ng)}{\mathcal{I}(x_0|\Nd)}=1+\kappa_\gamma\left(\frac{1+\Gedc/\Gebg}{1-D(x_0)\kappa_e}-1\right)
    \label{Eq:FGI}
\end{equation}
Notice this expression diverges for $\kappa_e=1$ and $D(x_0)\rightarrow1$. This is the same divergence we get by taking a ratio of the Bernoulli and Poisson FI: $\mathcal{I}_B/\mathcal{I}_P=1/1-\theta$ (see previous section).

\subsection{Estimating Sample Parameters in SEM}

\label{Sec:SEM_Est}

Using FI, we can estimate the SNR improvement for quantitative measurement in an SEM. To show this, we will model the tilt-dependent SE yield as 

\begin{equation}
    \delta(\theta)=\frac{\delta_0\delta_1}{\cos(\theta)(\delta_1-\delta_0)+\delta_0},
    \label{SEq:t2d}
\end{equation}

where $\theta$ is the surface tilt, $\delta_0$ is the SE yeild for a horizontal surface ($\theta=0$), and $\delta_1$ is the SE yield at the edge of a tall vertical step \cite{Postek2014}.  

In order to estimate the secondary electron yield $\delta$ and surface tilt $\theta$, we need to construct estimators $\hat{\delta}$ and $\hat{\theta}$ which are functions of the measurement data. The measurement data consists of tallies of the number of coincident events, $N_{e\gamma}$, electron-only events, $N_e$, and photon-only events, $N_\gamma$. For applied dose $d$, we use the estimators

\begin{align*}
    \hat{\delta}_h(N_{\gamma e},N_e,N_\gamma;d)&=\mathbb{E}_\delta\{p(\delta|N_{\gamma e},N_e,N_\gamma;d)\}\\
    \hat{\theta}_h(N_{\gamma e},N_e,N_\gamma;d)&=\mathbb{E}_\theta\{p(\theta|N_{\gamma e},N_e,N_\gamma;d)\}
\end{align*}
where $\mathbb{E}_X\{p(X)\}=\sum_x xp(X=x)$ is the expectation value of $X$. We give these estimators the $h$ subscript to differentiate them from the estimators used without heralding,
\begin{align*}
    \hat{\delta}(N_{\gamma e}+N_e;d)&=\mathbb{E}_\delta\{p(\delta|N_{\gamma e}+N_e,;d)\}\\
    \hat{\theta}(N_{\gamma e}+N_e;d)&=\mathbb{E}_\theta\{p(\theta|N_{\gamma e}+N_e;d)\}
\end{align*}
We can use Bayes' theorem to write
\begin{equation*}
    p(\delta|N_{\gamma e},N_e,N_\gamma;d)=\frac{p(\delta)}{p(N_{\gamma e},N_e,N_\gamma;d)}p(N_{\gamma e},N_e,N_\gamma|\delta;d)
\end{equation*}
where
\begin{equation*}
    p(N_{\gamma e},N_e,N_\gamma;d)=\int_{\delta_0}^{\delta_1}d\delta\ p(\delta)p(N_{\gamma e},N_e,N_\gamma|\delta;d)
\end{equation*}
and $p(\delta)$ is a probability distribution describing our expectations or prior knowledge of $\delta$. We use analogous expressions for $\theta$. For simplicity, we will assume $p(\theta)$ is uniform:
\begin{equation*}
p(\theta)=\begin{cases}\frac{2}{\pi}&0<\theta<\pi/2\\0&\text{else}
\end{cases}
\end{equation*}
The distribution $p(\delta)$ is induced by $p(\theta)$ and their relation in Eq. \ref{SEq:t2d}. Note, however, that uniform distributions are not always the best representation of maximum ignorance. 

The distribution $p(N_{\gamma e},N_e,N_\gamma|\theta;d)$ is a convolution of a Poisson distribution which determines the number of electrons produced by the gun and a multinomial distribution which selects the fate of each electron:
\begin{align*}
    &p(N_{\gamma e}=n_{\gamma e},N_e=n_e,N_\gamma=n_\gamma|\theta;d)=\\
    &\sum_{n_e^{(g)}}\mathcal{P}(n_e^{(g)};d)\mathcal{M}(n_{\gamma e},n_e,n_\gamma;n_e^{(g)},p_{\gamma e}(\theta),p_e(\theta),p_\gamma(\theta))
\end{align*}
where
\begin{equation*}
\mathcal{P}(n_e^{(g)};d)=\frac{e^{-d}d^{n_e^{(g)}}}{n_e^{(g)}!}
\end{equation*}
and
\begin{align*}
&\mathcal{M}(n_{\gamma e},n_e,n_\gamma;n_e^{(g)}(\theta),p_{\gamma e}(\theta),p_e(\theta),p_\gamma(\theta))=\\&\frac{n_e^{(g)}!}{n_{\gamma e}!n_e!n_\gamma!n_n!}p_{\gamma e}(\theta)^{n_{e\gamma}} p_e(\theta)^{n_e} p_\gamma(\theta)^{n_\gamma} p_n(\theta)^{n_n}
\end{align*}
Where $p_n=1-p_{\gamma e}-p_e-p_\gamma$ and $n_n=n_e^{(g)}-n_{\gamma e}-n_e-n_\gamma$, with the subscript $n$ indicating the scenario where an electron produced by the gun does not trigger the electron or photon detectors. 

The convolution can be simplified into a product of three Poisson distributions:
\begin{align*}
&p(N_{\gamma e}=n_{\gamma e},N_e=n_e,N_\gamma=n_\gamma|\theta;d)=\\&\mathcal{P}(n_{e\gamma}; p_{e\gamma}(\theta)d)\mathcal{P}(n_{e}; p_{e}(\theta)d)\mathcal{P}(n_{\gamma};p_{\gamma}(\theta)d)
\end{align*}
Finally, according to the model described above, we have
\begin{align*}
    p_{e\gamma}(\theta)&=g\eta_\gamma f_1D(\theta)\\
    p_{e}(\theta)&=(f-g\eta_\gamma f_1)D(\theta)\\
    p_{\gamma}(\theta)&=g\eta_\gamma(1-f_1D(\theta))
\end{align*}
with $D(\theta)=1-e^{-\eta_e\delta(\theta)}$.

\end{document}